\documentstyle[epsfig,12pt]{elsart}

\begin{document}
\newcommand{\p}{\partial}
\newcommand{\ls}{\left(}
\newcommand{\rs}{\right)}
\newcommand{\beq}{\begin{equation}}
\newcommand{\eeq}{\end{equation}}
\newcommand{\beqa}{\begin{eqnarray}}
\newcommand{\eeqa}{\end{eqnarray}}
\begin{frontmatter}
\title{Fragment Formation in Central Heavy Ion Collisions 
 at Relativistic Energies}
\author{E. Santini},
\author{T. Gaitanos},
\author{M. Colonna},
\author{M. Di Toro\thanksref{dit}}
\thanks[dit]{ditoro@lns.infn.it}

\address{Laboratori Nazionali del Sud INFN, Physics and Astronomy Department, 
University of Catania, I-95123 Catania, Italy } 
\begin{abstract}
We perform a systematic study of the fragmentation path of excited nuclear 
matter in central heavy ion collisions at the intermediate energy of 
$0.4~AGeV$. 
The theoretical calculations are based on a
Relativistic Boltzmann-Uehling-Uhlenbeck ($RBUU$) transport equation
including stochastic effects. A Relativistic Mean Field ($RMF$) approach
is used, based on a non-linear Lagrangian, with coupling constants
tuned to reproduce the high density results of calculations with
correlations. 
 
At variance with the case at Fermi energies, a new fast clusterization 
mechanism is revealed in the early compression stage of the reaction 
dynamics. Fragments appear directly produced from phase-space fluctuations
due to two-body correlations. In-medium effects of 
the elastic nucleon-nucleon cross sections on the fragmentation dynamics are 
particularly discussed. The subsequent evolution of the primordial clusters
 is treated using a simple phenomenological phase space coalescence 
algorithm. 

The reliability of the 
approach, formation and recognition, is investigated in detail by comparing 
fragment momentum space distributions {\it and simultaneously} their yields 
with recent 
experimental data of the $FOPI$ collaboration by varying the system 
size of the 
colliding system, i.e. its compressional energy (pressure, radial flow). 
We find an excellent agreement between theory and experiment in almost all the 
cases and, on the other hand, some limitations of the simple 
coalescence model. 
Furthermore, the temporal evolution of the fragment structure is explored 
with a clear evidence of an earlier formation of the heavier clusters, 
 that will appear as interesting $relics$ of the high density phase
of the nuclear Equation of State ($EoS$). 
\end{abstract}
\begin{keyword}
Heavy ion collisions at intermediate energies \sep
Fragment formation \sep Rapidity distributions 
\sep  Nuclear matter \sep Equation of state 
\sep In-medium cross sections 
\sep Dirac-Brueckner-Hartree-Fock
\PACS 25.75.-q, 25.75.Ld, 21.65.+f
\end{keyword}
\end{frontmatter}
\section{Introduction}
One of the major interests in the study of intermediate energy ($0.1-1$ AGeV) 
Heavy Ion Collisions ($HIC$) is the determination of the nuclear matter 
Equation of State ($EoS$) under conditions of density and/or temperature 
beyond 
saturation. During the last two decades many attempts have been successfully 
done in this direction, see Refs. \cite{ritter} for an overview. It turned out 
that baryon collective flow strongly depends on the high density behavior of 
the nuclear $EoS$. Experimental studies on collective flow have suggested a 
rather ``soft'' $EoS$ at supra-normal densities  \cite{dani} which 
has a similar 
functional dependence as that obtained from microscopic 
Dirac-Brueckner-Hartree-Fock 
($DBHF$) theory  \cite{DBT}. Also particle production, in particular 
subthreshold Kaon 
($K^+$) yields are affected by compressional effects of the high density 
region. 
Experiments on particle production have strongly supported a ``soft'' 
$EoS$ at high 
densities \cite{aich,kaons}.

However, one has to realize that a $HIC$ is a rather complicated 
non-equilibrium 
process. A unique determination of the nuclear $EoS$ far away from saturation 
requires a complete characterization of the collision dynamics in 
comparison with 
experimental data when available. It has been experimentally shown that the 
final state of a $HIC$ at intermediate energies is dominated by 
fragments with a strong collective flow pattern relative to that of free 
protons. In particular, more of $70\%$ of protons are bound to 
clusters \cite{fopi97}. 
The collective baryon flow is thereby connected to that of the fragment 
flow in terms of a linear dependence with respect to the fragment charge  
\cite{flowf1,flowf2}. It therefore turns out that the description of the 
process of 
fragmentation is very important in theoretical transport studies of an entire 
characterization of the reaction dynamics (apart the dynamical behavior of 
nucleons 
and produced particles). 

Here we will concentrate on the fragment production in central $HIC$
at intermediate energies. This study is particularly interesting since it 
could be
compared to the clusterization mechanism evidenced at lower energies,
 based on a growth of spinodal instabilities leading to a Liquid-Gas ($LG$)
phase transition \cite{ChomazPR389}. The latter  mechanism is active in 
the expansion
 phase of the excited nuclear system during the reaction dynamics 
 and so it gives information on the low density behavior of the nuclear
$EoS$. It is important to note that similar signatures have been found
in peripheral fragmentation at higher energies \cite{poch}, where actually
the first evidence of a nuclear $LG$ phase transition was revealed. 

Both scenarios, central collisions at the Fermi energies and projectile
fragmentation at intermediate energies, have in common the presence of a 
fragmenting source without a large radial flow, i.e. a relatively
low expansion velocity \cite{poch,essler}. In central collisions
in the $(0.1-1)AGeV$ range we have a much faster expansion of the
interacting nuclear matter and the spinodal mechanism will be
largely hindered due to the mismatch between the timescales of the
instability growth and of the expansion. The picture appears even worse
for a low-density nucleation mechanism. In fact in this case we have
a non-vanishing surface tension leading to a barrier which needs to
be overcome, with a further increase of the relevant time-scale.

Clusters are then expected to
come directly from correlations in the high density region, actually
partially reduced during the expansion. Consistently the size
distribution will be very different, much more dominated by light ions,
since the characteristic wavelengths of the mean field instabilities
 \cite{ChomazPR389} will not play any role. If this picture is correct,
 we will have a chance to see in the fragment properties some direct
effects of the nuclear $EoS$ at high density. This will be particularly
valid for the heavier fragments, that could be considered as the
$relics$ of the high density phase.

The microscopic transport models have proven to be an 
adequate tool for the 
description of the non-equilibrium reaction dynamics at intermediate energies 
\cite{danif,nantes,giessen,muenchen} (see for recent results \cite{v2fopi}). 
The physical input of such semi-classical models based on 
Boltzmann type equations are the nuclear mean field and the nucleon-nucleon 
cross sections. Both can be derived either directly from the effective 
two-body 
in-medium interaction, i.e. the in-medium G-matrix \cite{horror} or 
phenomenologically from 
Skyrme- and Walecka-type models \cite{blaettel,koli}. Although Boltzmann type 
transport 
models describe the dynamics of the single-particle distribution function very 
satisfactory, they do not provide information on the dynamical evolution of 
(physical) fluctuations which are important in understanding the fragmentation 
mechanism. 
Different ways have been proposed to include the evolution of higher 
order correlations beyond the mean 
field level: by adding a 
fluctuation term 
(Boltzmann-Langevin equation) \cite{bl}, by choosing the numerical 
fluctuations 
in a judicious way just by limiting the number of test particles \cite{fluc}, 
or finally by 
introducing fluctuations directly into the single-particle phase space 
distribution 
function \cite{ColonnaNPA642,colonna04}. The detailed consequences of 
these different approaches 
are still under intense investigation. Here we will follow the second 
procedure, numerically easy to implement in the transport code, based on the 
noise of a discrete mapping of the phase space. 
This approach allows to mantain in the dynamics the random effects 
of the collision term, that otherwise would be completely washed out
in presence of a large number of test-particles \cite{BertschPR160}.
Thus, the collision term will
be able to initiate fluctuations leading to a cluster formation, as
suggested in ref. \cite{BondorfPLB359}. This
description looks particularly suitable in the present case for clusters 
produced on a very short time scale in the compression phase.
Such high density clusterization mechanism has been suggested also in other
non-relativistic dynamical transport models, like Quantum Molecular Dynamics 
($QMD$) \cite{AichelinPR202,nantes,ZbiriIWM04} and 
Boltzmann-Uehling-Uhlenbeck ($BUU$) \cite{ChenPRC69} approaches. Here we
investigate in detail the fragment dynamics, yields and velocity 
distributions, and the dependence on the size of the colliding systems. 
 
The fragment recognition in the $HIC$ dynamics  
consists of a phenomenological procedure, the phase-space coalescence 
model \cite{coala}, 
which has been successfully implemented within mean field transport 
simulations in the 
past \cite{giessen,blaettel,gait99}. Its advantage is that it can be easily 
applied to the 
final phase space distribution function, in order to 
realistically 
compare with experiments which require information on fragment yields. 
The determination 
of the final state of a $HIC$ in terms of baryons {\it and} fragments is 
furthermore 
necessary in theoretical simulations in order to use  the same methods as 
in the experiments to characterize the events, i.e. 
reaction plane resolutions, centrality selections using charge particle 
multiplicities or 
other related observables. 
Here we extend its use even to  
 provide 
information on
how the system develops through after the initial cluster formation. 

 The idea of the present work is to investigate the 
reliability of this fragmentation path, formation and further evolution,
in intermediate energy heavy ion collisions, taking the chance
of the existence of new 
experimental 
data from the $FOPI$ collaboration 
\cite{fopinew1,fopinew2}. To do so, we first describe the main ingredients 
of the stochastic 
transport model and the phase space coalescence algorithm, and then apply them 
to $HIC$ reaction dynamics and compare the theoretical results with all 
available experimental data related to fragment production. In particular, the 
system size dependence of fragment velocity distributions and yields is 
investigated in detail.

\section{Fragment description in RBUU}

The traditional approach to theoretically investigate heavy ion collisions 
is the transport equation of a Boltzmann type, called as 
Relativistic-Boltzmann-Uhlenbeck-Uheling ($RBUU$) equation, which is 
 described in detail 
in refs. \cite{horror,blaettel,koli}. The $RBUU$ equation has the form 
\begin{eqnarray}
\Bigg ( k^{\ast}_{\mu} \partial^{\mu}_{x} + ( k^{\nu \ast} 
F^{\mu\nu} + m^{\ast}(\partial^{\mu} m^{\ast}) ) \partial_{\nu}^{k^{\ast}} 
\Bigg ) f(x,k^{\ast})  = \frac{1}{2} \frac{1}{(2 \pi)^3} 
\int \frac{d^3 k_1^{\ast}}{k_1^{\ast 0}} \frac{d^3 k^{\ast '}}{k_1^{\ast 0 '}} 
\frac{d^3 k_1^{\ast '}}{k_1^{\ast 0 '}} & & \nonumber\\
\times W(k^{\ast} k_1^{\ast}|k^{\ast '} k_1^{\ast '}) 
\Bigg [ f(x,k^{\ast '}) f(x,k_1^{\ast '}) (1 - f(x,k^{\ast}))
(1 - f(x,k^{\ast}_1)) & & \nonumber\\
- f(x,k^{\ast}) f(x,k_1^{\ast}) (1 - f(x,k^{\ast '}))
(1 - f(x,k^{\ast '}_1)) \Bigg ] & &
\label{RBUU}
\quad .
\end{eqnarray}
Eq. (\ref{RBUU}) describes the evolution of the single particle 
distribution function $f(x,k^*)$ 
under the influence of a mean field, which enters via effective masses $m^*$, 
 effective momenta $k^*$ and 
the field tensor $F_{\mu\nu}$, and of binary collisions. The numerical 
solution of the 
$RBUU$ equation is based on the test particle method using Gaussian functions 
(first introduced in Ref. \cite{nlv}), in which each nucleon is represented 
by a finite sum of test particles. 
In fact one uses covariant Gaussian's (in Minkowski and momentum 
space) adopting the Relativistic Landau-Vlasov ($RLV$) method \cite{rlv}. 
In \cite{essler} it was shown that the $RLV$ method is appropriate to produce 
smooth fields and it is possible to determine local quantities, such as 
densities, local momentum distributions, etc., without introducing additional 
grids.  
The collision integral includes all inelastic channels up to pion 
production. The energy and angular dependence of the inelastic cross 
sections are 
taken from Ref. \cite{crossin}. For the elastic cross sections we will use two 
options: (a) the free parametrizations according Cugnon et al. \cite{crossel} 
and (b) the in-medium effective cross sections from the Tuebingen-group 
\cite{crosseff}.

Within the Relativistic Mean Field ($RMF$) frame a non-linear Lagrangian is
used, which leads to
a soft $EoS$ at high densities, similar to that derived  from 
correlated Dirac-Brueckner-Hartree-Fock $DBHF$ approaches  
\cite{GrecoPLB562,GaitanosNPA732}.
In this way one is able to 
reproduce the single-particle dynamics very satisfactory in terms of 
collective 
flow observables of nucleons and other produced particles as outlined in the 
introduction \cite{GrecoPLB562,GaitanosNPA732}. 
In the isovector channel scalar and vector field contributions are
included, with coupling constants also derived from $DBHF$ estimations
\cite{GrecoPLB562,GaitanosNPA732}.
Moreover, a comparison with experiments requires in any case the 
knowledge of the 
degree of clusterization, in order to perform the flow analysis in the 
same way as 
in the experiment \cite{flowf1,flowf2}. 
 
According to the phase space coalescence procedure, a number 
of nucleons can form a cluster if their distances in coordinate and 
momentum space are smaller than 
a given set of coalescence parameters $R_{c},~P_{c}$, respectively. 
These parameters can be 
fitted by adjusting the charge distributions, as it has been already shown in 
Refs. \cite{gait99}. However, so far more detailed dynamical properties of 
fragments, recognized by means of a
phase space coalescence, have not yet been studied due to the missing of 
precise experimental information.

Very recently the $FOPI$ collaboration has performed a systematic analysis on 
fragment velocity distributions and yields in terms of rapidities 
 in beam and 
in transverse (with respect to the beam axis) directions 
\cite{fopinew1,fopinew2}. 
The experimental 
analysis has also been extended to study the system size dependence of the 
degree of clusterization and fragment flows. Thus, by fitting the coalescence 
parameters just on global charge distributions one is now able to test 
in detail the 
phase space 
coalescence model on more exclusive observables of the reaction dynamics in 
an essentially parameter-free way. 

\section{Application to heavy ion collisions}

\begin{figure}[t]
\begin{center}
\unitlength1cm
\begin{picture}(9,8.5)
\put(-2.25,0){\makebox{\epsfig{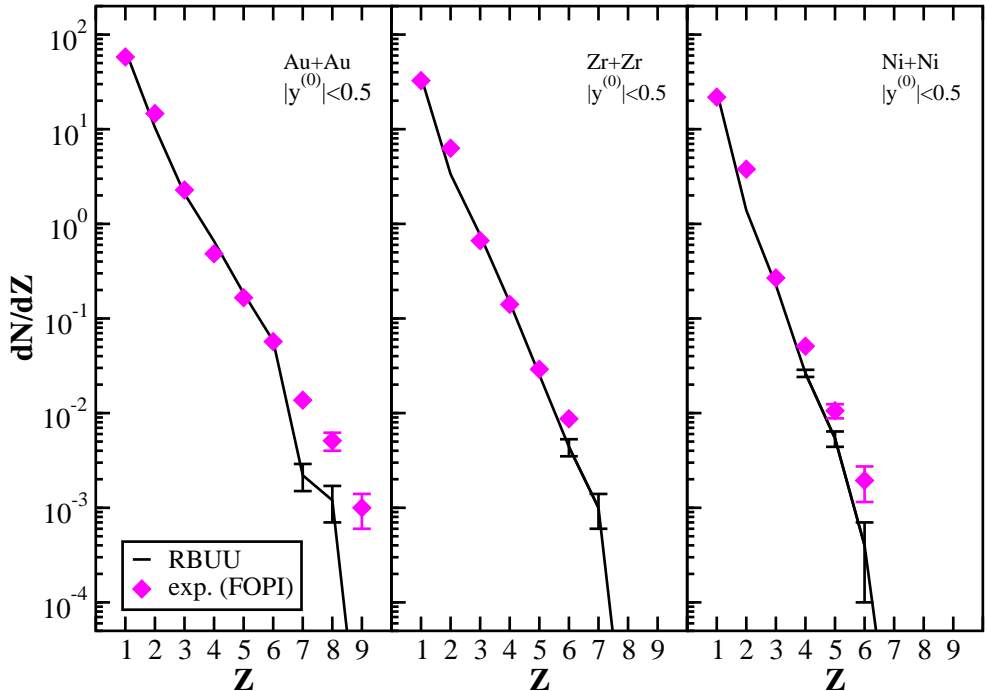}}}
\end{picture}
\caption{\label{Fig1} Charge distributions at mid rapidity ($|y^{0}|<0.5$,
 \protect\cite{redrap}) 
for central ($b^{(0)} \leq 0.15$, \protect\cite{redimp}) 
collisions for different systems (as indicated) at $0.4~AGeV$ incident energy. 
The theoretical curves (solid lines) performed within the $RBUU$ transport 
model 
are compared with experimental data (diamonds) from the $FOPI$ collaboration 
\protect\cite{fopinew1}.}
\end{center}
\end{figure}
We have performed simulations of heavy ion collisions at an intermediate 
incident energy per nucleon of $0.4~AGeV$ using the 
$RBUU$ equation for the evolution of the phase space distribution function and 
the phase space coalescence applied in the final state 
 of each $RBUU$ event. Clusters are identified at a $freeze-out$ time of
$90~fm/c$, when they are well separated in phase space. As shown later, when 
we will look at the time evolution of the cluster structures, these
 ``final'' results are not much depending on the choice of the 
$freeze-out$ time, which is obviously varying with the beam energy.
This is important in order to get information on system size effects
at a given beam energy.

In fact, in order to compare with the available data, the analysis has been 
performed by varying the system size, from $(Ca,Ca)$ to $(Au,Au)$
 reactions, and focussing on central collisions. Experimentally the 
observable $ERAT$ has 
been adopted to select the most central events. Theoretically we use the 
same observable (see Ref. \cite{muenchen}) with the result of a centrality 
resolution consistent with the experimental one. One should note 
that similar studies adopting the 
Isospin-Quantum-Molecular-Dynamics 
($IQMD$) model predict the same impact parameter selections \cite{flowf2}. 
All the results discussed in the following have been obtained 
with a soft $EoS$ (at supra-normal densities) with a compression modulus 
of $200~MeV$ and using $free$ and $effective$ $NN$ cross sections. 
In order to obtain a reasonable statistics for cluster studies, we 
have performed a coalescence procedure to $5000$ random samplings of $A$
nucleons for each
``stochastic'' $RBUU$ event, for given initial condition. 
As discussed before one ``stochastic'' $RBUU$ event corresponds to a transport 
calculation with about $50$ test particles (phase space gaussians) 
per nucleon. This number is in fact varying with the size of the system
in order to have the same global phase-space mapping when we change the number
of nucleons, a $total$ number of test particles around $15000$. We have 
checked that this mapping ensures a good time evolution of mean one-body
observables allowing the development of local fluctuations from 
direct nucleon-nucleon interactions. 

\subsection{Charge particle distributions}

Fig. \ref{Fig1} shows the charge particle distributions at mid rapidity 
for three different 
colliding systems ($Au+Au$, $Zr+Zr$ and $Ni+Ni$ at $0.4~AGeV$ 
 incident energy). 
It can be seen that the theoretical results fit the experimental data very 
well. The extracted coalescence parameters have been chosen as $4.5~fm$ and 
$1.5~1/fm$ 
in coordinate and momentum space, respectively. One should note that these 
parameters, once fixed from the $Au+Au$ charge distribution, are unique for 
all the systems considered here. 

In fact these estimations represent some ``effective'' coalescence parameters,
 partially related to the amount of fluctuations inserted in the $RBUU$
simulations via the discrete test-particle mapping of phase-space. Indeed from
 the sampling procedure described before we can expect that if we increase the
total number of test-particles we can reduce the coalescence parameters.
As already noted we have to find a compromise with the need of a limited
number of test-particles that will allow the development of local
fluctuations. We have checked the stability of the results vs. a factor 
two change of the total number of test-particles given before. It is essential
to keep the coalescence parameters fixed for all studied colliding systems,
 we can then trust the reliability of the physics results.

The moderate 
discrepancies between theory and experiment for the heavier clusters 
($Z > 5-6$), reported in Fig. \ref{Fig1}, will not influence most of the 
results 
shown below, due to their 
low multiplicity with respect to that of protons and light ions.
We will see later that in fact these heavier fragments are associated
to a very interesting physics and so the lack of statistics, in 
simulations as well as in experiments, represents a serious drawback.
 
We note how well the simulations are reproducing the increasing slope
of the $N(Z)$ exponential behavior with decreasing system size.
This doesn't mean that lighter systems become ``hotter'' but just
that we have less stopping while the fragments, in particular the heavier
ones, are produced in the dense phase. This ``non-equilibrium'' interpretation
will be clear in the following and in fact it has been already suggested
from the analysis of experimental data in ref. \cite{fopinew1}.

In connection to the previous point, with the fixed coalescence 
parameters one can now start to study the 
reaction dynamics in terms of the degree of transparency and related 
variances, i.e. in terms of the rapidity distributions of fragments 
along the beam and transverse directions. 

\begin{figure}[t]
\begin{center}
\unitlength1cm
\begin{picture}(9,8.5)
\put(-2.25,0){\makebox{\epsfig{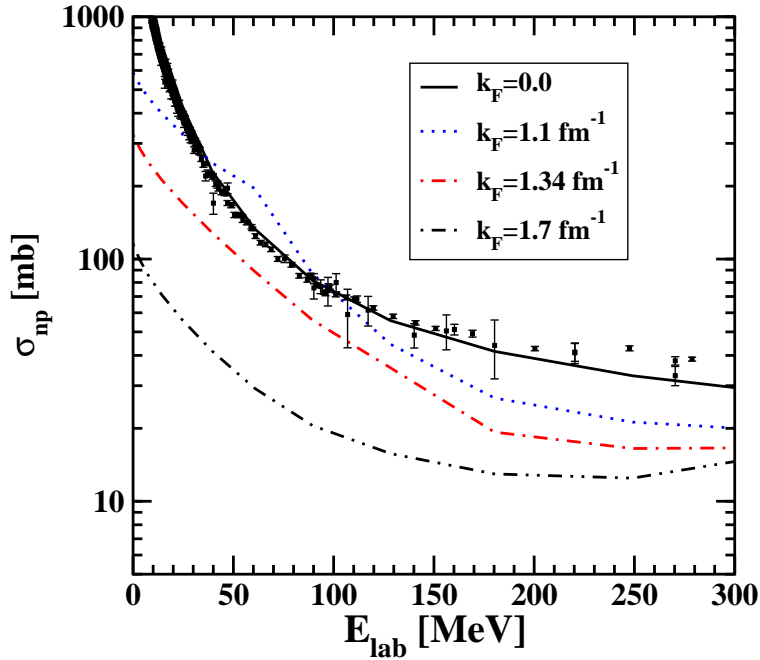}}}
\end{picture}
\caption{\label{Fig2} Elastic in-medium neutron-proton cross sections 
at various Fermi momenta 
$k_{F}$ as function of the laboratory energy $E_{lab}$. The free cross
section ($k_{F}=0$) is compared to the experimental total $np$ cross
section \protect\cite{pdg}(crosses).}
\end{center}
\end{figure}
In central collisions at intermediate energies the degree of stopping is 
mainly influenced by the binary collisions or the viscous behavior of 
the system. More (less) collisions lead to less (more) viscosity, i.e.
more (less) local equilibration and 
thus to less (more) transparency of the colliding matter. The connection 
between the reaction dynamics and the viscous behavior has been studied in 
Ref.  \cite{fopinew2,gait04b}. It was shown that in-medium effects of the 
elastic 
nucleon-nucleon ($NN$) cross sections are important for a reliable 
description of the 
reaction dynamics as far as the baryon stopping and flow is concerned. 
It is therefore of great interest 
to see whether in-medium effects of the $NN$ cross section influence the 
fragment dynamics. We will study first this important topic in the following 
subsection.

\subsection{In-Medium effects of the elastic NN cross sections on clusters}

Fig. \ref{Fig2}  shows the energy dependence of the in-medium 
neutron-proton $(np)$ cross section \cite{crosseff} at 
Fermi momenta $k_F = 0.0, 1.1,1.34,1.7 fm^{-1}$, 
corresponding to the densities 
 $\rho \sim 0,0.5,1,2\rho_0$ ($\rho_0=0.16 fm^{-3}$ 
is the nuclear matter saturation density). 
\begin{figure}[t]
\begin{center}
\unitlength1cm
\begin{picture}(9,8.5)
\put(-2.25,0){\makebox{\epsfig{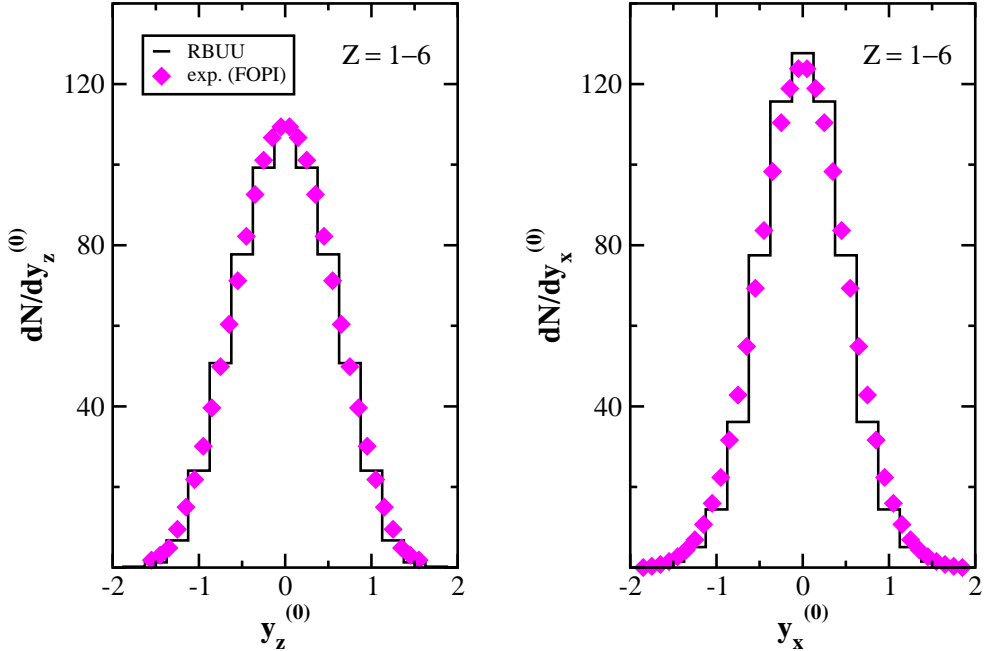}}}
\end{picture}
\caption{\label{Fig3}Longitudinal (left) and transversal (right) rapidity 
distributions \protect\cite{redrap} for all charged ions up 
to $Z=6$ (weighted with their 
charge) for central ($b^{(0)} \leq 0.15$) $Au+Au$ collisions at $0.4~AGeV$ 
incident energy. Theoretical $RBUU$ calculations (solid histograms) are 
compared with $FOPI$ data (diamonds) from \protect\cite{fopinew2}.}
\end{center}
\end{figure}
The presence of the medium leads to a substantial 
suppression of the cross section which is most pronounced at 
low laboratory energy $E_{\rm lab}$ and high densities where, in addition to 
the $(m^*/m)^2$ scaling, the Pauli-blocking of intermediate states is 
most efficient \cite{crosseff}. At larger $E_{\rm lab}$ asymptotic values of 
$15-20~mb$ are reached. However, not only the total cross sections 
but also the angular distributions are affected by the presence of the 
medium. The initially highly forward-backward 
peaked  $(n,p)$ cross sections becomes much more isotropic at finite densities 
\cite{crosseff} which is mainly do to the Pauli suppression of 
soft modes ($\pi$-exchange) and corresponding higher partial waves in 
the T-matrix. In Ref. \cite{gait04b} it was shown that the reduced effective 
cross sections considerably influence the degree of transparency and improves 
the comparison with the data. This is illustrated in Fig. \ref{Fig3}
for the $Au+Au$ case,  
where the longitudinal and transversal rapidity distributions of all 
charged particles (weighted with their charge) are displayed. The 
theoretical calculations using the effective cross sections reproduce the 
experimental data very well. Not only the longitudinal, but also 
the transversal 
rapidity distribution compares well with the experimental data. The same 
conclusion is valid even for other incident energies, for details 
see Ref. \cite{gait04b}. 

\begin{figure}[t]
\begin{center}
\unitlength1cm
\begin{picture}(9,8.5)
\put(-2.25,0){\makebox{\epsfig{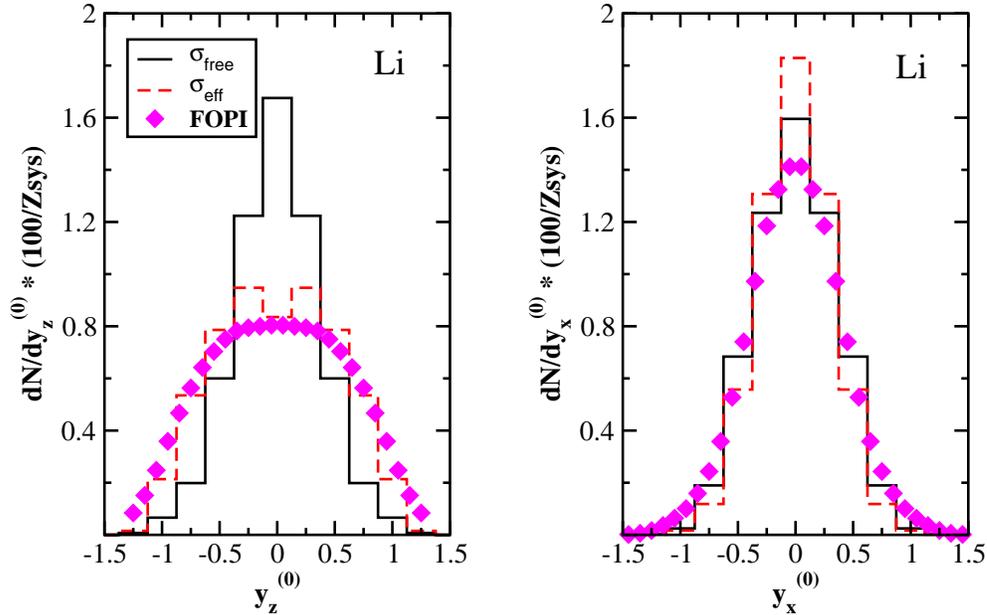}}}
\end{picture}
\caption{\label{Fig4} Scaled longitudinal (left) and transversal (right)
 rapidity 
distributions for $Li$-fragments for central ($b^{(0)} \leq 0.15$) 
$Ru+Ru$ collisions at $0.4~AGeV$ incident energy. Theoretical calculations 
using free and in-medium $NN$ cross sections (solid and dashed histograms, 
respectively) are compared with $FOPI$ data 
(diamonds, from \protect\cite{fopinew1}). The ordinates are 
normalized to a constant system size of $Z=100$ nuclear charge, as in ref.
\protect\cite{fopinew1}.}
\end{center}
\end{figure}

Obviously the question appears if the fragment dynamics is affected by 
density effects on the $NN$ cross sections. 
The in-medium effects of the microscopic $NN$ cross sections influence 
the stopping features not only of the protons but particularly those of 
the fragments, as shown in Fig. \ref{Fig4} for the $Ru+Ru$ case. Here the 
longitudinal and 
transverse rapidity distributions of $Z=3$ clusters obtained from transport 
calculations using free and the effective $NN$ cross sections are 
compared with 
each other and with recent data. It is shown that the transparency effect is 
more pronounced with the (reduced) in-medium $NN$ cross sections and the 
distributions fit 
better with the experiment in this case. 

This observation, which is similar to 
that of protons, is a non-trivial feature. The in-medium effects on the cross 
sections become important at higher densities due to the influence of the 
(intermediate state) Pauli operator and the reduction of the effective mass. 
Thus this result clearly indicates that fragments are formed earlier,
 during the  stage where the local densities are still high, then one can 
obviously 
expect in-medium effects on the stopping power of fragments, and also
on the yields (total areas in Fig.\ref{Fig4}). 
In any case it turns out that the use of in-medium effects within a consistent 
basis (mean field and collision integral) is crucial in describing the 
reaction 
dynamics of central collisions. 
In conclusion all the simulation results shown here have been
obtained using the in-medium parametrizations of the $NN$ cross sections
discussed before, with the related energy, density and angular dependences
 \cite{crosseff}.
\subsection{Transparency features of protons and clusters}

\begin{figure}[t]
\begin{center}
\unitlength1cm
\begin{picture}(9,8.5)
\put(-2.25,0){\makebox{\epsfig{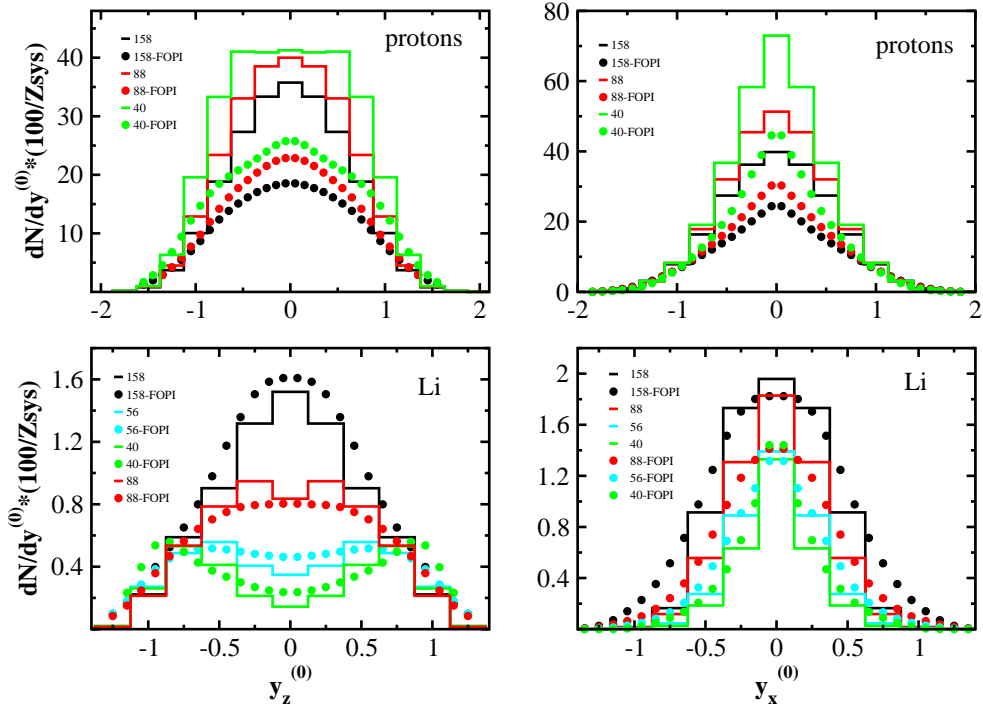}}}
\end{picture}
\caption{\label{Fig5}Scaled longitudinal (left) and transversal (right) 
rapidity 
distributions of single protons (top) and $Li$ clusters (bottom) in central 
($b^{(0)} \leq 0.15$) collisions at $0.4~AGeV$ incident energy for different 
symmetric systems. The various system charges $Z_{sys}$ are indicated in 
the figure. The ordinates are normalized to a common reference system 
with charge $Z_{sys}=100$. Theoretical calculations (histograms) are compared 
with $FOPI$ data (symbols) from \protect\cite{fopinew1}.}
\end{center}
\end{figure}

Fig. \ref{Fig5} shows the longitudinal ($dN/dy_{z}$) and transversal 
($dN/dy_{x}$) 
rapidity distributions of free protons (top) and light clusters (bottom) 
for central collisions for different colliding systems, as indicated. 
The centrality classes ($b^{(0)} \leq 0.15$) has been extracted in the same 
way as in the experiment \cite{fopinew1}. In general we observe a 
degree of transparency since the longitudinal rapidity distributions are 
in all 
cases broader compared to the transverse ones. 
However the size dependence of the transparency 
effect is indicatively different for protons vs fragments, see the left panels
of Fig. \ref{Fig5}. For $(Z=3)$ ions
it increases 
as the system size decreases: for the lightest system ($Ca+Ca$) the 
fragments are mainly formed in the spectator regions, lower-left panel
of Fig. \ref{Fig5}. 
The transparency effect of protons shows just the opposite trend.
These results, in nice agreement with the data, further support
the interpretation of an important dynamical cluster formation at higher 
densities. For light systems we cannot build high densities at mid-rapidity
and we observe a sharp drop in the fragment yield to the advantage of
a proton emission. In conclusion the  internal composition of the source 
is different for 
participant ($\vert y_z^{(0)} \vert <0.5$) and spectator 
($\vert y_z^{(0)} \vert \sim 1$) matter with respect 
to the system size of the colliding system. The heavier system exhibits a 
stronger ``liquefaction'' whereas by decreasing the size one observes 
an essential reduction of clusterization of the mid rapidity source. 
Only close to the spectator regions one sees an ``universality'' behavior 
independent of the system size 
(apart the lightest $Ca$-system) exactly like in the data. 
This universality behavior should not be 
confused with similar findings of the $ALADIN$ collaboration where peripheral 
collisions were studied with a clear separation of spectator fragmentation 
\cite{aladin}, see the detailed discussion in ref.\cite{fopinew1}.

The transverse rapidity distributions, right panels of Fig. \ref{Fig5},
appear also very instructive. As in the data, the proton yields for heavier
systems are dropping in the smaller $\vert y_x^{(0)} \vert <0.7$ transverse
rapidities, where more clusters are formed. The widths of the cluster
distributions are systematically increasing with the size, in the simulations
as well as in the data. In a statistical picture this could indicate 
an increase of the source temperature with the size, at variance with
the indications of the charge particle distributions, see the comments after
Fig.\ref{Fig1}. In a dynamical interpretation of the fragment production
the apparent contradiction disappears. The reduced stopping of the lighter 
systems, responsible of the faster decrease of the $N(Z)$ curves in 
Fig.\ref{Fig1}, will also imply reduced fluctuations in the nucleon phase
space distributions. Moreover, we will have even a reduced radial flow
that is decreasing the transverse velocity widths, as stressed
in ref.\cite{fopinew1}. In conclusion fragments are more formed if matter is
more stopped. This evidence is further supporting the ``high-density''
origin. In fact at the Fermi energies we observe just the opposite:
the multiplicity of fragments, produced now in the low-density phase,
is decreasing when increasing the $NN$ cross sections \cite{baranNPA730}.

 A difficulty appears in the absolute values of the (free) proton yields,
 upper panels of Fig.\ref{Fig5}.
 On the other hand, our estimations were appropriate 
for the multiplicities of $Z=1$ ions, Fig.\ref{Fig1} and the (weighted) 
rapidity distributions for $Z=1-6$ charges, Fig.\ref{Fig3}. 
A possible explanation could be the failure 
of the present naive
phase space coalescence model in describing the deuteron multiplicities. 
One can think that the description of all the fragments with the same 
set of coalescence parameters would not be appropriate for deuterons due to 
their relatively large $rms$ radius ($r_{d}=1.96$ fm \cite{deut}, e.g.
compared to the 
radius of $t$ and $He$ fragments, $r_{t,He}=1.61,1.74$ fm \cite{heli}, 
respectively). 
In fact, we obtain more free protons relative to deuterons. This explains 
the fact that the $Z=1$ multiplicity (which does not distinguish between 
protons and $Z=1$ fragments) can be well reproduced, in contrast to the 
free protons multiplicities. This point, that could be improved, in fact
is not modifying the physics of fragment production.

\begin{figure}[t]
\begin{center}
\unitlength1cm
\begin{picture}(9,8.5)
\put(-2.25,0){\makebox{\epsfig{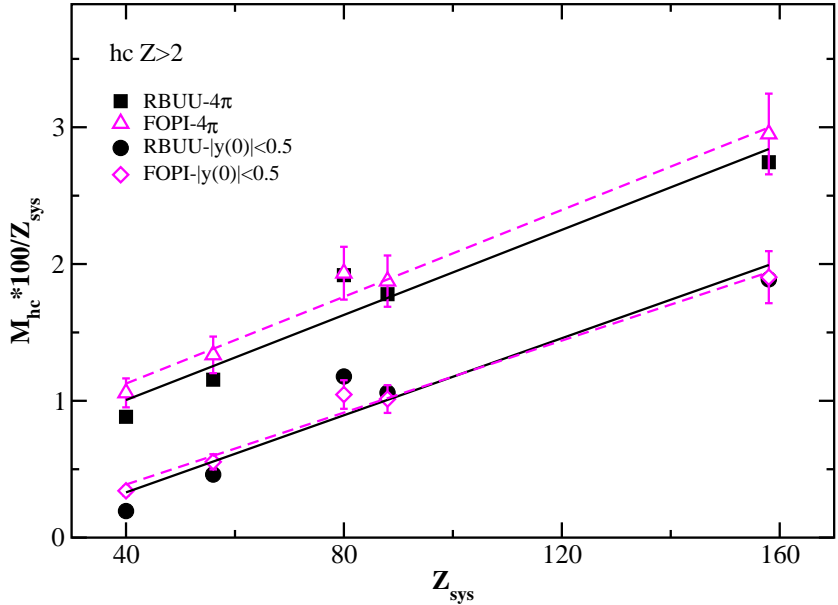}}}
\end{picture}
\caption{\label{Fig6}Average multiplicities of heavy ($Z > 2$) clusters, 
 normalized to those of a system with $100$ protons 
(i.e., $M_{hc} \times 100/Z_{sys}$), as function 
of the system charge Z. Calculations (full symbols) and experimental data 
(empty) for all 
the phase space ($4\pi$) and confined to the midrapidity interval 
$|y_z^{(0)}|<0.5$ are shown. All straight lines are linear least square 
fits to the 
symbols. The $FOPI$ data are taken from Ref. \protect\cite{fopinew1}.
}
\end{center}
\end{figure}

\begin{figure}[t]
\begin{center}
\unitlength1cm
\begin{picture}(9,8.5)
\put(-4.5,9.5){\makebox{\epsfig{file=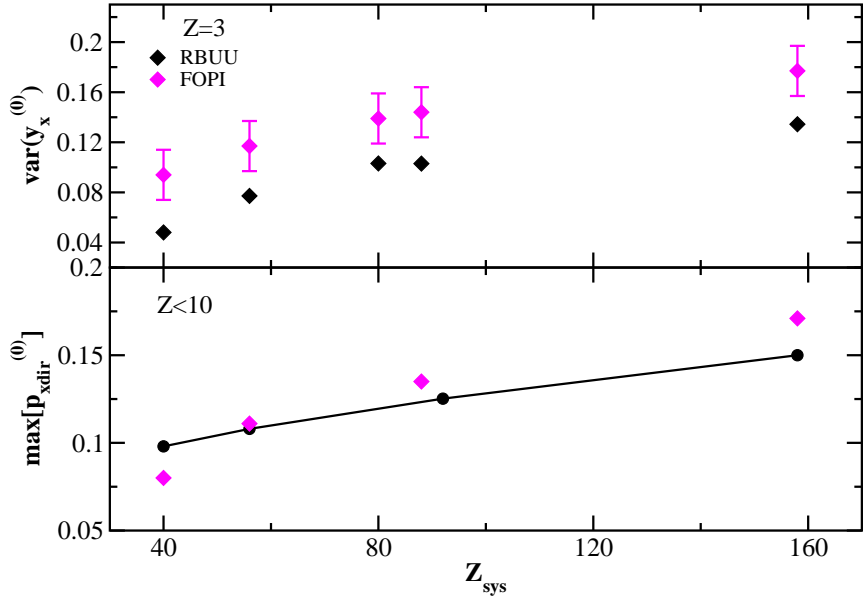,width=10.0cm}}}
\end{picture}
\caption{\label{Fig7} Size dependences of: i) Variances of the transverse 
rapidity distributions for Li (upper panel), central 
($b^{(0)} \leq 0.15$) collisions, data from 
\protect\cite{fopinew1}; ii) Maximal global sideflow
(lower panel), semicentral collisions ($b^{(0)} \simeq 0.4$),
 data from \protect\cite{fopinew2}. The symbols have the 
same meaning as 
in the previous Fig. \protect\ref{Fig6}.
}
\end{center}
\end{figure}
The system size dependences of the fragment multiplicities, degree
of stopping and maximum sideflow  
are summarized in Figs.\ref{Fig6},\ref{Fig7} and compared to the
corresponding $FOPI$ data. 

The multiplicity 
of heavy clusters $Z>2$, $M_{hc}$ (relative to that of a 100 proton system), 
linearly increases 
with system size, independent of the phase space selections 
(apart the absolute 
values). The comparison between the theoretical calculations and the data 
is almost perfect, see Fig.\ref{Fig6}. In particular, the multiplicity of 
heavier clusters is well reproduced quantitatively, including the
deviation in the system size systematics for $Z \approx 80$. 
This refers to the $Ru,Zr$ systems (same mass number $A$, but different 
isospin content) and
may be due to isospin effects. However, it is a very moderate effect 
and will not 
be discussed in the following. As already noted in the data, 
ref.\cite{fopinew1}, the good linear fits mean that the system-size
dependence of $M_{hc}$ follows a quadratic behavior vs. the system
charge (mass). This further supports the two-body dominant mechanism
for fragment formation.

In Fig.\ref{Fig7} we compare to the data the size dependence of
some global features of the reaction dynamics, stopping (upper panel)
and maximum directed flow (lower panel). This is important to check
if our transport simulations are simultaneously well reproducing
attractive (radial-flows) and repulsive (side-flows) observables.
For the stopping we report the variances in the transverse rapidity
distributions of $Li$ ions in central collisions.
The scaled directed flow is defined as in \cite{fopinew2},
 $p_{xdir}^{(0)} \equiv p_{xdir}/u_{proj}$, where 
$p_{xdir} = \Sigma sign(y) Z u_x/ \Sigma Z$ ($Z$ fragment charge,
$u_{proj}$ spatial part of the projectile 4-velocity, 
$u_x \equiv \beta_x \gamma$ projection of the fragment 4-velocity on the
reaction plane). The sum is over all charged ions
with $Z<10$ and $y$ the related rapidity. The maximum values reported
in the simulation points of the Fig.\ref{Fig7} (lower) correspond
to the $b^{(0)} = 0.4$ scaled impact parameter \cite{redimp}
collision for the various systems, see \cite{fopinew2}.

The agreement is satisfying. The stopping is increasing with the 
system size, 
indicated in the enhanced values of the transverse variance with $Z$ in Fig. 
\ref{Fig7}(upper panel). The transport estimations are systematically  a little
below the data since the tails of the distributions are underestimated,
see the Fig.\ref{Fig5} bottom-right panel. This could indicate a 
slightly reduced radial flow. The agreement is better for the side-flow,
Fig.\ref{Fig7}(lower panel).

This good agreement for fragment velocity distributions is not obvious, since 
the phenomenological parameters of 
the phase space coalescence model were {\it globally} fixed to charge 
distributions, 
without taking care of the momentum distributions. In this context it is 
worthwhile noting that a corresponding analysis in the framework of the 
Isospin Quantum-Molecular Dynamics ($IQMD$) model shows the same stopping
results but cannot reproduce the 
fragment multiplicities with such an accuracy \cite{fopinew1}. 

In general we can state that once we have fixed the physical 
parameters of the transport descriptions, i.e. the density dependence of the 
nuclear $EoS$ and the in-medium $NN$ cross sections, it turns out 
that the rather 
simple phase space coalescence model for fragment recognition 
works astonishingly well at intermediate 
incident energies. This is an important conclusion in view of the fact that 
most of 
transport models based on $(R)BUU-type$ approaches make use of phase space 
coalescence 
to simulate experimental selections and to reconstruct the centrality 
classes in the 
same way as in experiments, etc. \cite{flowf2}. We have now to better
analyze the cluster formation mechanism before trying to extract some
physics information on the nuclear $EoS$. 

\section{Clusters as probes of the nuclear EoS}

The simulation $and$ experimental results discussed in the previous
section are confirming the expectations that in central
$HIC$ at intermediate energies fragments are originally formed
in the high density phase of the reaction dynamics from fluctuations
due to two-body correlations. This interpretation is mostly supported
by two clear evidences: i) The correlation between fragment multiplicity
and stopping, see Fig.\ref{Fig6} and Fig.\ref{Fig7}(upper panel); ii) The
linear $Z_{sys}$ behavior of the normalized heavy cluster multiplicity
of Fig.\ref{Fig6}. Such fast clusterization mechanism is completely
different from the one observed at the Fermi energies, the growth of
spinodal instabilities in dilute matter leading to a first order
liquid-gas phase transition.

This new fragmentation dynamics opens the exciting possibility of directly
probing the high density features of nuclear matter from the study
of the cluster properties. However, the high density formed clusters
would be largely modified during the expansion phase up to the freeze-out
time. So it appears very important to follow the dynamical evolution of the
clusterization in order to select the fragments that are expected to better
keep the memory of the primordial high density source.

\begin{figure}[t]
\begin{center}
\unitlength1cm
\begin{picture}(9,8.5)
\put(-2.25,0){\makebox{\epsfig{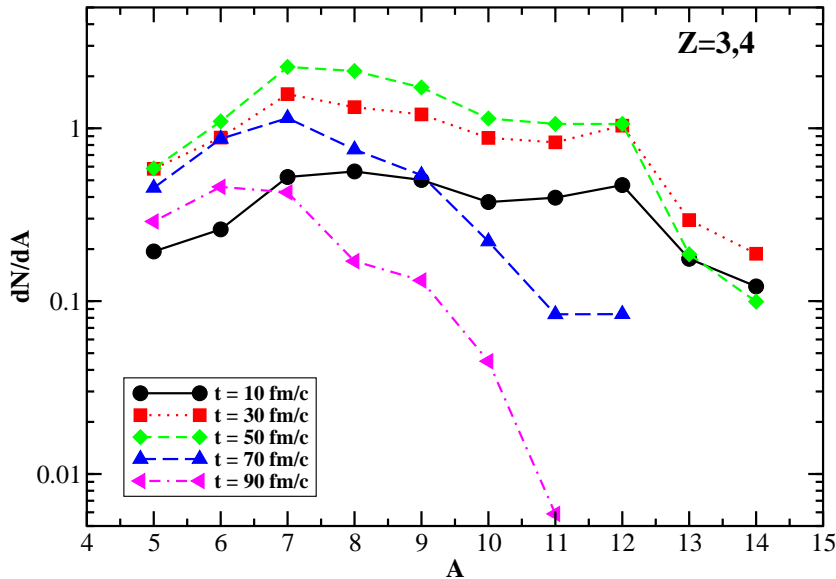}}}
\end{picture}
\caption{\label{Fig8} 
Mass distributions of clusters with charges in the interval $Z=3,4$ at
 different 
times, as indicated, of a $Au+Au$ central collision at $0.4~AGeV$ 
incident energy.
}
\end{center}
\end{figure}
We have applied the phase space 
coalescence at several stages of a central $Au+Au$ collision. 
Fig.\ref{Fig8} shows 
the mass distributions of 
$Z=3,4$ fragments at different times as indicated. It is seen 
that heavy 
clusters are identified at a very early phase of the reaction dynamics 
($t \approx 10-30$ fm/c). In this time interval the system is in a
high density phase, of around $2.5\rho_0$ and locally equilibrated,
so we can speak in terms of the nuclear $EoS$ 
 \cite{GrecoPLB562,GaitanosNPA732}.
After that time-step the system enters a fast expansion phase and the 
multiplicity 
of heavy clusters 
starts to decrease with a corresponding enhancement of the number 
of light (stable) $Li$ and $Be$ isotopes. We note 
that no sequential decay has been applied so far.

It turns out that heavy clusters are formed very early during the high density 
phase whereas the evaporation of single nucleons and light fragments appear at 
stages after freeze-out. This picture seems not obvious since 
at that early stage the colliding system is still hot and highly excited. 
A possible answer for this surprising result could be the onset of a 
collective motion responsible for a fast cooling of the fireball region. 
In fact, in theoretical as well as in experimental studies,  a 
strong (isotropic) radial collective flow pattern is found in intermediate 
energy collisions 
of heavy nuclei such as $Au$ \cite{giessen,gait00,eos}. The radial flow 
sets in very early 
during the expansion phase and then rapidly governs the dynamics. Thus 
much of thermal energy is transfered into collective motion making the 
existence of heavy primary fragments possible. However, the radial flow 
pattern 
does not ``freeze'' the multiplicity of the early formed heavy clusters. 
In fact the momentum distribution of nucleons inside those clusters is still 
rapidly changing because of 
the strong radial flow component (apart the relatively small thermal and Fermi 
momentum components). Thus, due to the coalescence requirement in momentum 
space, the radial flow will imply
 an effective break-up of primary heavy fragments. 
This mechanism finally leads to the formation of light clusters. 

This scenario seems to be not unrealistic: radial flow is ultimately connected 
with pressure gradients and thus with the achieved maximal densities. With 
respect to the system size the maximal density is (almost) proportional to the 
number of $participant$ nucleons as the degree of stopping does, 
\cite{fopinew2}. Similar behavior will show up in the $NN$ collision 
frequency. Therefore, 
heavy clusters are 
preferentially formed in the fireball region between heavy colliding systems 
relative to that of the lighter ones. As already noted, this interpretation 
can explain the rapidity 
distributions and the fragment multiplicities of the previous section
(in particular the quadratic $Z_{sys}$ dependence of Fig.\ref{Fig6}).

We conclude that the phase space 
coalescence model can successfully characterize the non-trivial reaction 
dynamics of heavy ion collisions at intermediate energies. Furthermore, 
the interesting evidence of an early heavy-cluster formation offers the 
possibility 
to study compressional and isospin effects at the level of fragmentation
measurements. 
Studies based on the $IQMD$ transport model \cite{nantes} confirms indeed 
an early cluster formation and a moderate $EoS$ (isoscalar) dependence of 
the fragment multiplicities \cite{fopinew2}. Thus, it would be interesting 
to explore whether 
also high density isospin effects can be studied in terms of fragmentation 
dynamics at relativistic energies.

\section{Summary and conclusions}

We study in detail the mechanism of fragment production in
central collisions at intermediate energies (at $0.4~AGeV$). In particular
we investigate the size dependence of the process in collisions of
ions of different masses, where nice recent data are existing.
We present a rather complete comparison of theoretical results on 
velocity distributions {\it and simultaneously} on multiplicities 
of all the produced particles. 

Using a stochastic transport model we show the evidence of a new fast
clusterization mechanism, present in the early compression stage of the
reaction dynamics. The formed fragments are then propagating through the
expansion phase up to the freeze-out, underlying subsequent breakings
mostly due to a dynamical effect of the radial flow. As a consequence the
$survived$ heavier fragments will represent the $relics$ of the high density 
phase and then could be used as direct probes of such hot and dense
nuclear matter.

This scenario is supported by a series of quantitative observations, in
full agreement with the existing data:
\begin{itemize}
\item {The size dependence of charge distributions at mid-rapidity:
 heavier clusters are more produced from heavier systems, where larger
desisies can be reached in the compression stage.}
\item{The clear correlation between heavy cluster ($hc$, $Z>2$) multiplicity
and global stopping.}
\item{The quadratic dependence of cluster multiplicities on masses (charges)
of the colliding ions, nice indication of the link between the $seeds$
of the clusterization and the two-body correlations.}
\end{itemize}

An important support to this interpretation is the possibility of a
simultaneous reproduction of other global observables of the
fragment dynamics, like the directed flows. The fragment recognition
in all these transport simulations has been based on the very simple
phase-space coalescence algorithm. Such approach appears rather reliable
and this is important in view of further comparisons with experimental data,
 in particular for the choice of the same event selections. In fact we
have also seen some limits, i.e. in the evaluation of deuteron vs. proton
yields, but this point could be properly improved, see \cite{ChenPRC69}.

The conclusion is that observables related to fragment production
in central collisions at intermediate energies can provide new
independent information on the nuclear $EoS$ at supra-normal densities.
In particular the measurement of the isospin content of the heavier fragments
appears very interesting. In the $Au+Au$ case the presence of a
$Isospin~Distillation$, i.e. more protons bound in the clusters, would be
a nice indication of a $stiff$ symmetry term well above saturation density,
 of large astrophysics interest. This could be a good motivation for
fragmentation studies at $SIS$ energies with radioactive beams.





\begin{thebibliography}{00}
\bibitem{redrap} $y_{z,x}^{(0)}$ are the $reduced$ (longitudinal/transverse)
 rapidities normalized to the Projectile ones.
\bibitem{redimp} $b^{(0)}$ is the $reduced$ impact parameter, normalized
to a $b_{max}=1.15(A_p^{1/3}+A_t^{1/3})$.
\end{thebibliography}
\end{document}